\documentclass[twocolumn,showpacs,amsmath,amssymb,prb,superscriptaddress,a4paper]{revtex4}

  \usepackage[T1]{fontenc}       
  \usepackage[utf8]{inputenc}
  \usepackage[frenchb, english]{babel} 
  \usepackage{amsmath}
  \usepackage{xcolor} 
  \usepackage{graphicx}  
  \usepackage{graphics}          
  \usepackage{dcolumn}
  \usepackage{pgf,pgfarrows,pgfautomata,pgfheaps,pgfnodes,pgfshade}
  \usepackage{multimedia}
  \usepackage{braket}
  \usepackage{lmodern}
  \usepackage{picture}
  \usepackage{psfrag,fancyhdr,layout,appendix,subfigure}
  \usepackage{epstopdf}
  \usepackage{placeins}
  \usepackage{bm}
  \usepackage{anysize}
  \usepackage{times,color}
  \usepackage{subfigure}
  \usepackage{mathptmx}
  \usepackage{amssymb}
  \usepackage{amsfonts}
  \usepackage{amsmath}

\DeclareGraphicsExtensions{.jpg,.mps,.pdf,.png,.gif}

\begin{document}

\bibliographystyle{plain}

\preprint{APS/123-QED}

\author{Catherine Tonin}
\email{tonin@insp.jussieu.fr}

\affiliation{Institut des NanoSciences de Paris, Universit\'{e}
Pierre et Marie Curie, CNRS UMR 7588, 4 Place Jussieu, 75005 Paris, France}

\author{Richard Hostein}

\affiliation{Institut des NanoSciences de Paris, Universit\'{e}
Pierre et Marie Curie, CNRS UMR 7588, 4 Place Jussieu, 75005 Paris, France}

\author{Valia Voliotis}

\affiliation{Institut des NanoSciences de Paris, Universit\'{e}
Pierre et Marie Curie, CNRS UMR 7588, 4 Place Jussieu, 75005 Paris, France}

\author{Roger Grousson}

\affiliation{Institut des NanoSciences de Paris, Universit\'{e}
Pierre et Marie Curie, CNRS UMR 7588, 4 Place Jussieu, 75005 Paris, France}

\author{\\Aristide Lema\^{\i}tre}

\affiliation{Laboratoire de Photonique et Nanostructures, CNRS,
Route de Nozay, 91460 Marcoussis, France}

\author{Anthony Martinez}

\affiliation{Laboratoire de Photonique et Nanostructures, CNRS,
Route de Nozay, 91460 Marcoussis, France}

\title{Polarization properties of excitonic qu-bits in single self-assembled quantum dots}

%\date{\today}
\date{}

%%%%%%%%%%%%%%%%%%%%%%%%%%%%%%%%%%%%%%%%%%%%%%

\begin{abstract}

We investigate polarization properties of neutral exciton emission in single self-assembled InAs/GaAs quantum dots. The in-plane shape and strain anisotropy strongly couple the heavy and light hole states and lead to large optical anisotropy with non-orthogonal linearly polarized states misaligned with respect to the crystallographic axes. Owing to a waveguiding experimental configuration, luminescence polarization along the growth axis has been observed revealing the presence of shear components of the deformation tensor out of the growth plane. Resonant luminescence experiments allowed determining the oscillator strength ratio of the two exciton eigenstates. Valence band mixing governs this ratio and can be very different from dot to dot, however the polarization anisotropy axis is quite fixed inside a scanned area of one $\mu m ^2$ and indicates that the in-plane deformation direction to which it is related has a correlation length of the order of magnitude of one $\mu m ^2$.

\end{abstract}

%%%%%%%%%%%%%%%%%%%%%%%%%%%%%%%%%%%%%%%%%%%%%%

\pacs{71.35.-y, 78.47.Fg, 78.67.Hc}
\maketitle

%%%%%%%%%%%%%%%%%%%%%%%%%%%%%%%%%%%%%%%%%%%%%%

%%%%%%%%%%%%%%%%%%%%%%%%%%%%%%%%%%%%%%%%%%%%%%

%%%%%%%%%%%%%%%%%%%%%%%%%%%%%%%%%%%%%%%%%%%%%%

%%%%\section{Introduction}

%%%%%%%%%%%%%%%%%%%%%%%%%%%%%%%%%%%%%%%%%%%%%%

%%%%%%%%%%%%%%%%%%%%%%%%%%%%%%%%%%%%%%%%%%%%%%

\section{Introduction}

Addressing and manipulating an individual spin in a single quantum dot (QD) is a fundamental step to achieve for eventual information processing based on quantum mechanics. \cite{Imamoglu, Xu, Ramsay} Neutral excitons can also be used as single qu-bits but their polarization properties have to be precisely known in order to address a specific state. It is well established now that for self-assembled quantum dots, in-plane shape anisotropy induced during growth leads to two linearly polarized orthogonal exciton states, splitted by the exciton fine structure splitting (FSS),  \cite{Gammon, Bayer, Goupalov, Bester, Karlsson10, Flissikowski, Favero, Astakhov} having in principle the same oscillator strength. 

However, recent experimental observations show that the linear states are neither aligned along the crystallographic axes, nor orthogonal between them and an important linear polarization degree can be found. \cite{Favero} These facts strongly indicate that valence-band mixing (VBM) due to the anisotropic confinement potential and strain effects should be taken into account to explain the strong polarization anisotropy observed in the emission of individual QDs, as well as the exciton eigenstates orientation.

Indeed, in II-VI CdTe/ZnTe \cite{Koudinov, Leger} or III-V InAlAs/AlGaAs \cite{Ohno} QDs, VBM has been related to the strain anisotropy induced during growth process. In strain-free GaAs QDs grown by droplet epitaxy it has been shown that VBM exists due to the anisotropy of the confinement potential,  \cite{Belhadj} while in strained self-assembled InAs/GaAs QDs, VBM effects have been investigated in order to estimate the exciton spin relaxation time, \cite{Kowalik} its influence on the exciton FSS, \cite{Plumhof, Lin} or the hyperfine coupling. \cite{Testelin} To our knowledge, no experimental study includes both strain effects and confinement anisotropy in order to explain the observed results, which can be fully described by combining the Luttinger-Kohn and the Bir-Pikus Hamiltonians. \cite{LK,BP}

We report on polarization resolved micro-photoluminescence ($\mu PL$) of the neutral exciton emission in single self-assembled InAs/GaAs QDs. We investigate the role of VBM on polarization properties, due to confinement potential anisotropy and local strain effects described by a Luttinger-Kohn and Bir-Pikus Hamiltonian. 

Since strain effects and dot shape anisotropy depend strongly on growth conditions they vary from sample to sample and from dot to dot. Therefore, a large dispersion of values for the heavy to light-hole mixing parameters can be found in the literature. \cite{Ohno, Leger, Belhadj, Lin} Moreover, since morphological details and precise values for the different strain components can not be obtained at the same time for each studied QD by optical experiments, it is quite impossible to know precisely the origin of valence-band mixing. 

Nevertheless, from the study of polarization anisotropy on a very large statistical ensemble of individual QDs (about 300 dots on two different samples), we can conclude that: (i) the polarization anisotropy axis direction is dominated by strain-induced VBM, (ii) the polarization main axis is not aligned along the crystallographic axes but has a given direction within one probed micron, and (iii) the VBM strength can be large and can vary significantly from dot to dot in the same one probed micron. These results seem to indicate that if the polarization main axis is related to the in-plane main strain orientation,\cite{Ohno, Leger} the mean components of the strain tensor are quite homogeneous for all dots inside a $1 \mu m^2$ area. However the strength of the mixing seems to be mostly related to confinement anisotropy.

Using an original experimental configuration where the QDs layer is embedded in a one-dimensional waveguide, resonant emission of the neutral exciton in single QDs could be analyzed and yield the two eigenstates oscillator strength ratio. Using this geometry, polarization-resolved $\mu PL$ analyzed along the growth axis could also be recorded, revealing coupling between same-spin heavy and light hole states due to off-plane strain components. In conventional geometries this can not be evidenced and only in-plane strain components are usually taken into account. \cite{Ohno, Leger}

The paper is organized as follows: in section II we give some details on the samples and the experimental configuration. Section III is devoted to the in-plane polarization – resolved $\mu PL$ experiments on single QDs excited non-resonantly (part A) and resonantly (part B). In parallel, simulations have been performed in order to give the general trends of the polarization properties influenced by mixing effects. In order to deduce the amplitude of the hh-lh coupling and the anisotropy axis of the polarization the experimental results are fitted taking into account the VBM. Moreover, an analysis of the eigenstates polarization misalignment with respect to the crystallographic axes is made. Section IV presents the results of polarization resolved $\mu PL$ along the z-axis which allowed to evidence the shear strain components out of the growth plane. 

%%%%%%%%%%%%%%%%%%%%%%%%%%%%%%%%%%%%%%%%%%%%%%

%%%%%%%%%%%%%%%%%%%%%%%%%%%%%%%%%%%%%%%%%%%%%%

%%%%\section{Samples and experimental set-up}

%%%%%%%%%%%%%%%%%%%%%%%%%%%%%%%%%%%%%%%%%%%%%%

%%%%%%%%%%%%%%%%%%%%%%%%%%%%%%%%%%%%%%%%%%%%%%

\section{Samples and experimental set-up}

\begin{figure}
      \includegraphics[width=7cm,height=4cm]{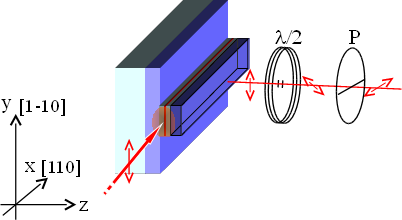}
\caption{(Color on-line) Waveguiding configuration. The laser propagates along the x-axis  and is polarized along y. The $\mu PL$
is detected along the z-axis and the polarization is analyzed by a half-wave plate followed by a polarizer. The WL and the
 QDs are represented as a red plane embedded in a waveguide formed by AlGaAs/ GaAs layers in blue and grey 
respectively. One micron ridges are etched on the sample surface to form the one-dimensional monomode optical guide.}
\label{expolar}
\end{figure}

InAs/GaAs self-assembled QDs, were grown by MBE on a planar $[001]$ GaAs substrate. In order to address the fundamental exciton state at resonance, the dots are embedded in a one-dimensional GaAlAs monomode waveguide (WG). \cite{Melet2} The wave-guiding geometry has several advantages: first, due to the spatial confinement of light, the volume of the optical mode is reduced and the coupling with the QDs is enhanced. Second, contrary to the microcavity case, there is no need to match the cavity mode energy with the QD emission energy in order to achieve a strong coupling between light and matter. Finally, the laser light propagation axis can be perpendicular to the luminescence detection direction. Indeed, while the laser propagates in the (x,y) dots plane, the single QD luminescence can be collected from the WG top surface by a confocal $\mu PL$ detection set-up  (see Fig. \ref{expolar}). Since the laser beam is confined in the guided mode, the scattered light is greatly suppressed and the resonant luminescence is almost background-free, at low pump power. Using this geometrical configuration, resonant Rabi oscillations of the exciton state in a single QD have been observed. \cite{Melet} Coherent control experiments have also been performed in order to determine the main decoherence mechanisms in the system. \cite{Enderlin} Moreover, in this geometry, it is also possible to excite a dot by the sample top surface and to detect the PL at the edge of the WG. This last configuration allows us to investigate the PL properties along the z-axis, which in usual back scattering geometry can not be resolved.

The QDs are excited by ps pulses provided by a mode-locked Ti:sapphire laser and focused on the sample using a microscope objective. The PL signal is then collected by another objective ($N.A. = 0.5$) and detected with a 0.5 m grating spectrometer and a liquid-nitrogen-cooled charge-coupled device camera. All the objectives are mounted on piezo-electric actuators so they can easily control the coupling to the WG and move along the ridge to probe different QDs.

The spatial resolution of the set-up is diffraction-limited to $1 \mu m^2$, while its spectral resolution is about $45\mu eV$. In order to resolve the neutral exciton FSS, a Fabry-Perot interferometer, used as interferential filter can be inserted in the collection path, improving our spectral resolution to $0.6 \mu eV$. Then we were able to measure the exciton FSS $\delta_{FSS}$, which varies depending on the dot from 20 to 145 $\mu eV$. The laser is polarized along $[1\overline{1}0]$  and the sample is fixed on the cold finger of a three-axes optical cryostat (10K) specifically designed for the wave-guiding geometry.

Two experimental configurations can be used for the polarization-resolved experiments: either the pump laser is guided and the $\mu PL$ is collected from the top surface of the WG, or the dot is excited from the top surface and its PL is collected at the edge of the WG. In both cases, the PL polarization is analyzed using a rotating half-wave plate (HWP) and a fixed linear polarizer, in order to avoid selection from the polarization response of the whole detection setup.

%%%%%%%%%%%%%%%%%%%%%%%%%%%%%%%%%%%%%%%%%%%%%%

%%%%%%%%%%%%%%%%%%%%%%%%%%%%%%%%%%%%%%%%%%%%%%

%%%%\section{Valence-band mixing in the dots plane (xy)}

%%%%%%%%%%%%%%%%%%%%%%%%%%%%%%%%%%%%%%%%%%%%%%

%%%%%%%%%%%%%%%%%%%%%%%%%%%%%%%%%%%%%%%%%%%%%%

\section{Valence-band mixing effects in the (x,y) dots plane}

In the case of QDs with revolution symmetry around the growth direction z ($[001]$ axis), the two neutral exciton bright eigenstates $\ket{+1}$ and $\ket{-1}$, with a total angular momentum $J = \pm 1$, are degenerated in energy and their emission is circularly polarized, left  ($\sigma^+$) or right ($\sigma^-$) in the dots plane (x,y). When the system symmetry is reduced, the electron-hole exchange interaction lifts this degeneracy and the two bright eigenstates are the two linear combinations \cite{Bester, Goupalov} $\ket{X} = \frac{1}{i\sqrt{2}}(\ket{+1} - \ket{-1})$ and $\ket{Y} = \frac{1}{\sqrt{2}}(\ket{+1}+\ket{-1})$, whose emission is linearly polarized along  the crystallographic axes [110] (x-axis) and [1$\overline{1}$0] (y-axis). \cite{Gammon, Bayer, Favero, Goupalov, Bester, Karlsson10} These two eigenstates are split in energy by the FSS ($\delta_{FSS} = |E_X-E_Y|$), and under non-resonant excitation the emission intensity of both states is the same, whatever the difference in oscillator strength. \cite{Favero}

If we now consider valence-band mixing (VBM) between heavy hole (HH) and light hole (LH) states which is due in general to in-plane shape asymmetry and anisotropic strain effects, the neutral exciton eigenstates become linear combinations of the elliptically-polarized bright exciton states $\ket{\pm\tilde{1}}$, which can be written as:
\begin{equation}
\ket{\pm\tilde{1}} = \sqrt{1-\beta^2} \ket{\mp\frac{1}{2};\pm\frac{3}{2}}+ \beta e^{\pm2i\psi} \ket{\mp\frac{1}{2};\mp\frac{1}{2}}
\end{equation}
The first spin state corresponds to the electron spin projection, while the second corresponds to the heavy ($\pm3/2$) and light ($\pm1/2$) hole spin projection. The simulation adjustable parameters $\beta$ and $\psi$ represent the amplitude and phase of the mixing as defined in the Luttinger-Kohn and Bir-Pikus (LKBP) Hamiltonian (see appendix \ref{appendix2}). Emission of these two states can then be elliptically polarized and their main axis can be tilted by an angle $\psi$ with respect to the y-axis [1$\overline{1}$0].
The two neutral exciton eigenstates are then:
\begin{equation}
\ket{\tilde{X}} = \frac{1}{i\sqrt{2}}(\ket{+\tilde{1}} - e^{2i\theta}\ket{-\tilde{1}})
\end{equation}
\begin{equation}
\ket{\tilde{Y}} = \frac{1}{\sqrt{2}}(\ket{+\tilde{1}} +e^{2i\theta} \ket{-\tilde{1}})
\end{equation}
An additional angle $\theta$ (see Fig. \ref{cahuetpouet}) has been introduced to take into account the QD main elongation axis orientation with respect to [1$\overline{1}$0]. As discussed in Refs. \citenum{Leger} and \citenum{Ohno}, this angle can be different from the crystallographic axes and $\psi$.

Depending on the LH ratio $\beta$ and the angle $\psi$, the two eigenstates $\ket{\tilde{X}}$ and $\ket{\tilde{Y}}$ can be tilted with respect to the crystallographic axes [110] and [1$\overline{1}$0], without being perpendicular anymore. The tilt angles are defined as $\phi_{\tilde{X}}$ and $\phi_{\tilde{Y}}$ respectively (see Fig \ref{phi}) and the angle between them is $\Omega$. In addition, under non-resonant excitation, because of the VBM, the two states have no longer the same emission intensity, resulting in the blue and red curves represented in Fig \ref{phi}.

\subsection{Non-resonant excitation}

\begin{figure}
\begin{center}
\hspace{0.9cm}
\subfigure[]{
      \includegraphics[width=0.2\textwidth]{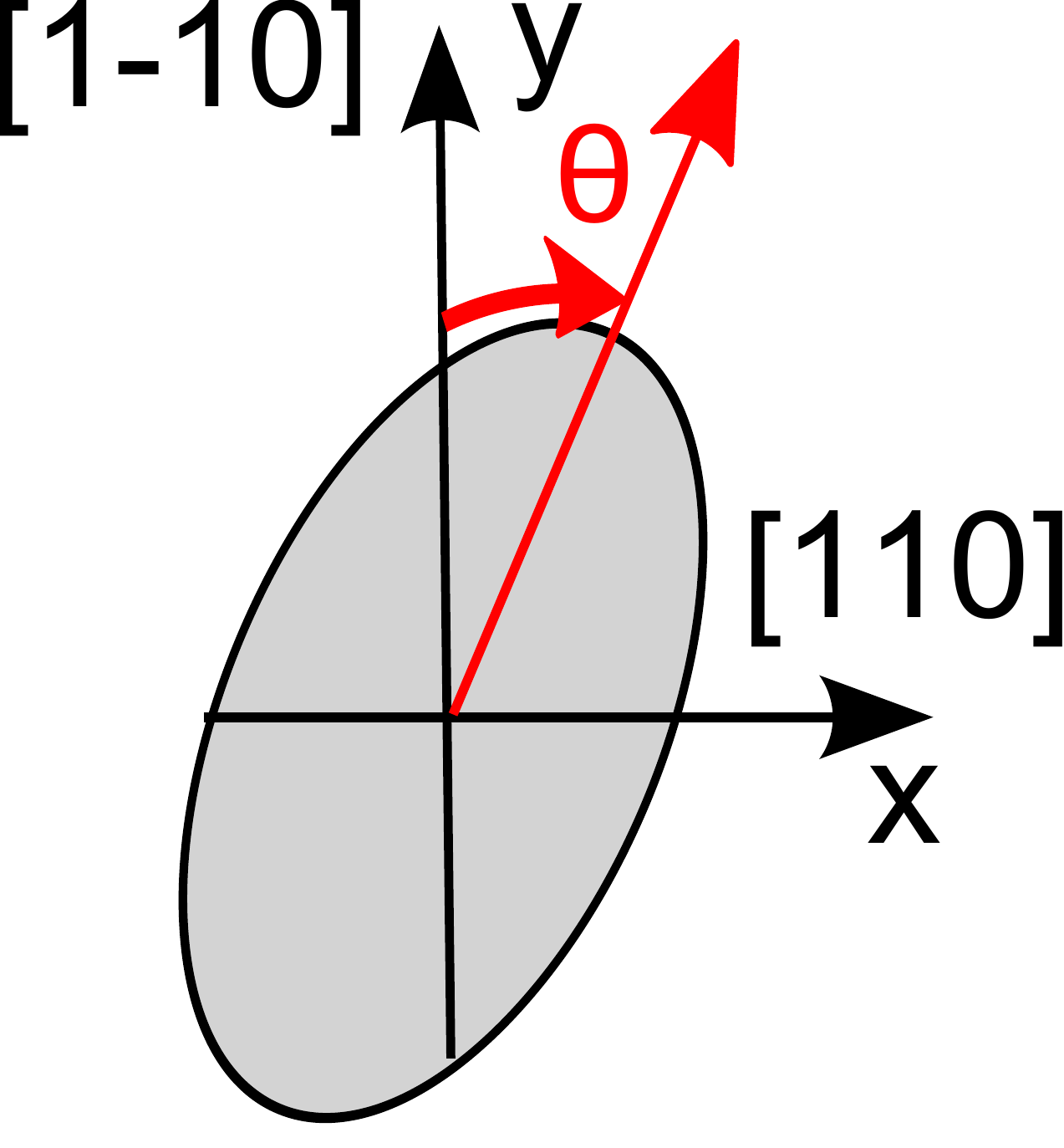}
\label{theta}}
\end{center}
   \begin{minipage}[c]{0.45\linewidth}
\subfigure[]{
      \includegraphics[width=1.2\textwidth]{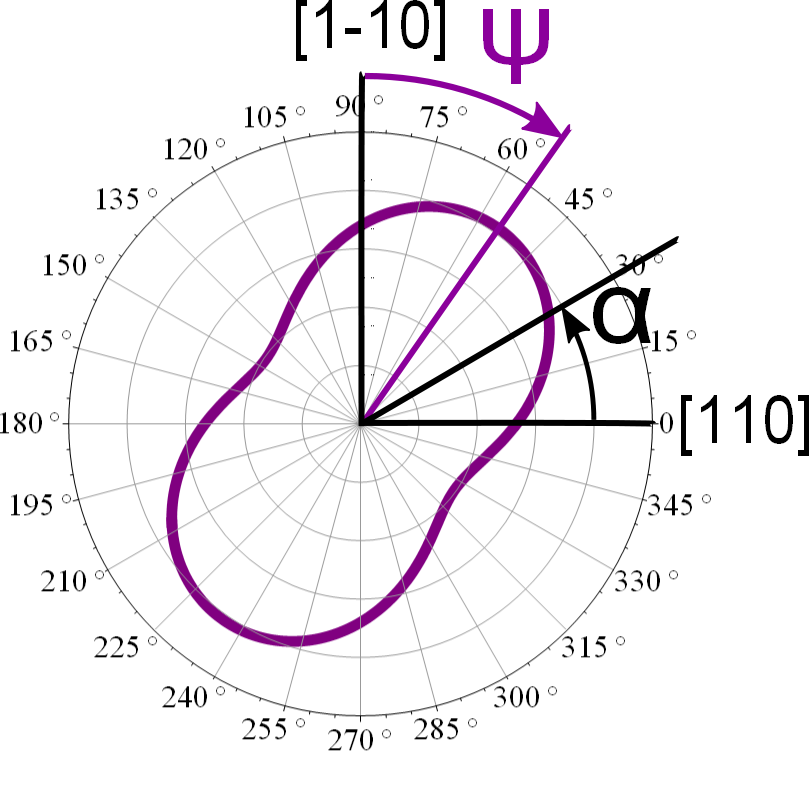}
\label{psi0}}
   \end{minipage} \hfill
   \begin{minipage}[c]{0.45\linewidth}
\subfigure[]{
      \includegraphics[width=1.1\textwidth]{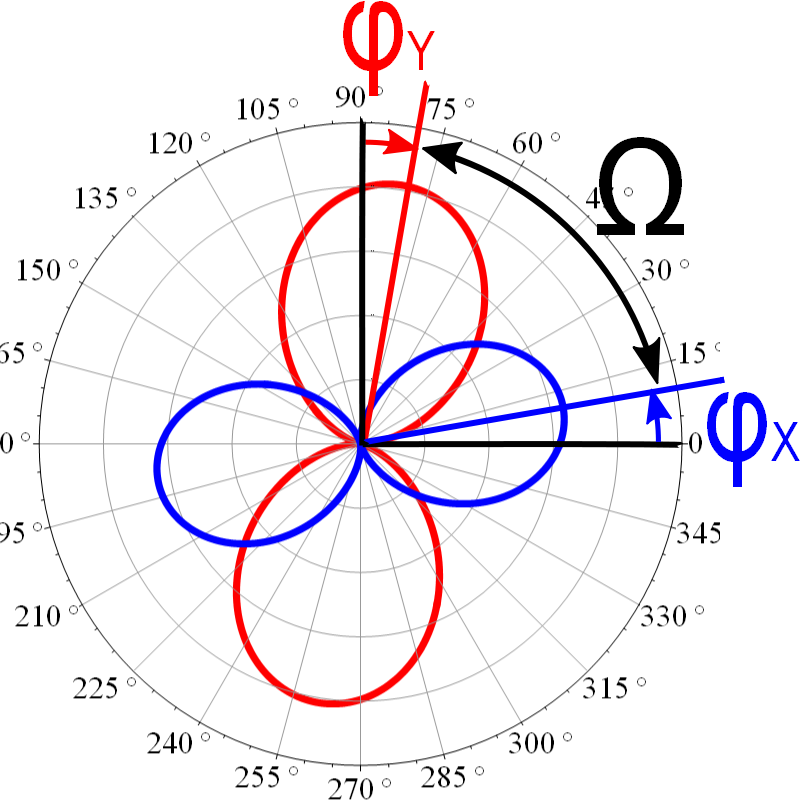}
\label{phi}}
   \end{minipage} \hfill
\caption{Definition of the different angles involved in the simulations. (a) $\theta$ is the main dot orientation axis. (b) $\psi$ is the main emission anisotropy axis and $\alpha$ is the angle between the analyzer axis and [110]. (c) $\phi_{\tilde{X}}$ and $\phi_{\tilde{Y}}$ are the eigenstates tilt angles and  $\Omega$ is the angle between the two $\ket{\tilde{X}}$ and $\ket{\tilde{Y}}$ eigenstates.}
\label{cahuetpouet}
\end{figure}

By calculating the dipole matrix elements for the transitions between the $\Gamma_6$ conduction-band and the $\Gamma_8$ valence-band states (see appendix \ref{appendix3}), the two eigenstates emission intensity, passing through a linear polarizer with an angle $\alpha$ with respect to the x-axis [110] (see Fig. \ref{psi0}) can be determined. Under non-resonant excitation, assuming that the two levels are equally populated, independently of the laser polarization and their respective oscillator strength, we obtain the normalized intensity:
\begin{eqnarray}
I_{\tilde{X}}(\alpha) & = & [ \sqrt{1-\beta^2} \cos(\alpha+\theta)  \nonumber\\
 & & - \frac{\beta}{\sqrt{3}} \cos(\alpha -\theta+ 2\psi)]^2 
\label{int1}
\end{eqnarray}
\begin{eqnarray}
I_{\tilde{Y}}(\alpha) & = & [ \sqrt{1-\beta^2}\sin(\alpha+\theta)  \nonumber\\
 & & + \frac{\beta}{\sqrt{3}} \sin(\alpha -\theta+ 2\psi)]^2
\label{int2}
\end{eqnarray}
The total intensity in the (x,y) plan, which off-resonance is simply the sum of the two expressions above, is then written:
\begin{eqnarray}
  \lefteqn{I_{nr}(\alpha) = [ 1-\frac{2}{3} \beta^2} \nonumber\\
 & & -2 \beta  \sqrt{\frac{1-\beta^2}{3}}\cos(2(\alpha -\psi))] 
\label{int tot}
\end{eqnarray}

\begin{figure}
\begin{center}
\hspace{-0.6cm}$\theta = 0 \degres$
\end{center}
   \begin{minipage}[b]{0.2\linewidth}
\subfigure[$\beta = 0$]{
      \includegraphics[width=3.5cm,height=3.5cm]{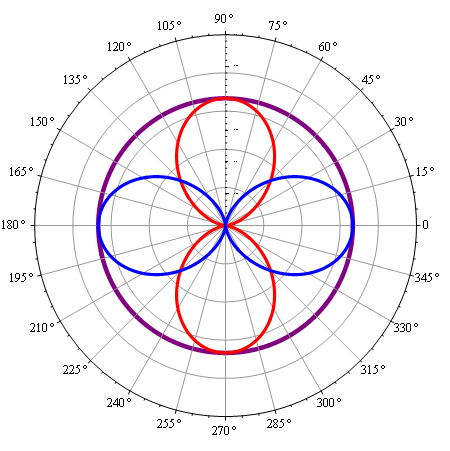}
\label{simu1a}}
   \end{minipage} \hfill
   \begin{minipage}[b]{0.6\linewidth}
\subfigure[$\beta = 0.25$; $\psi = 0\degres$]{
      \includegraphics[width=3.5cm,height=3.5cm]{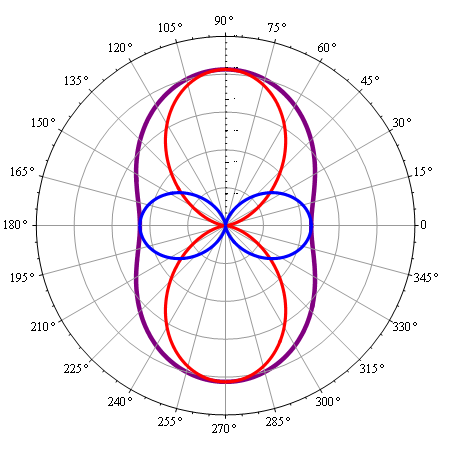}
\label{simu1b}}
   \end{minipage}
   \begin{minipage}[b]{0.2\linewidth}
\subfigure[$\beta = 0.25$; $\psi = 45\degres$]{
      \includegraphics[width=3.5cm,height=3.5cm]{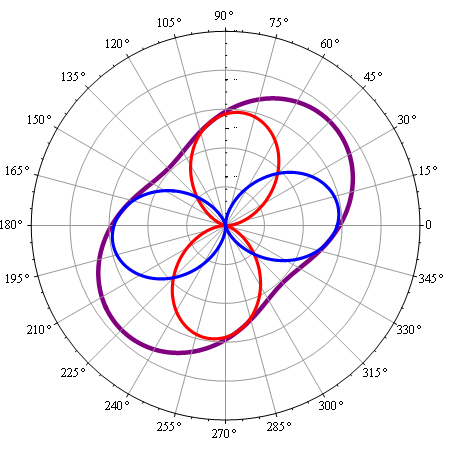}
\label{simu1c}}
   \end{minipage} \hfill
   \begin{minipage}[b]{0.6\linewidth}
\subfigure[$\beta = 0.25$; $\psi = 90\degres$]{
      \includegraphics[width=3.5cm,height=3.5cm]{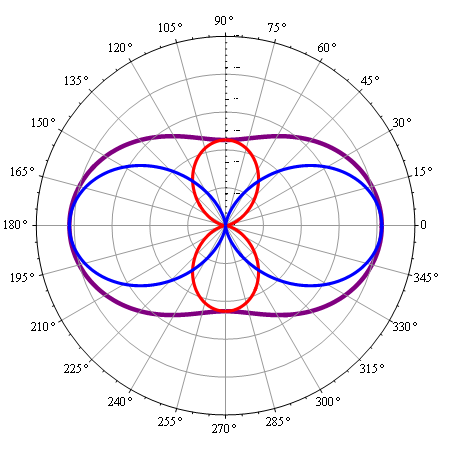}
\label{simu1d}}
   \end{minipage}
\vspace{0.5cm}
\begin{center}
\hspace{-0.5cm}$\theta = +10 \degres$
\end{center}
   \begin{minipage}[b]{0.2\linewidth}
\subfigure[$\beta = 0$]{
      \includegraphics[width=3.5cm,height=3.5cm]{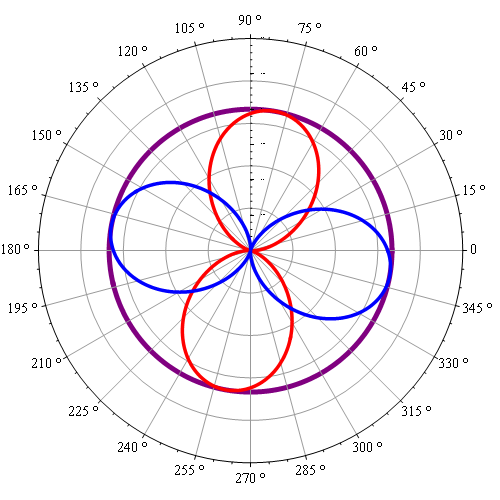}
\label{simu1g}}
   \end{minipage} \hfill
   \begin{minipage}[b]{0.6\linewidth}
\subfigure[$\beta = 0.25$; $\psi = 0\degres$]{
      \includegraphics[width=3.5cm,height=3.5cm]{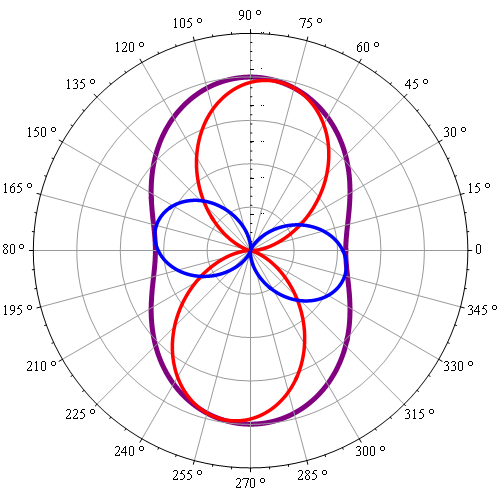}
\label{simu1h}}
   \end{minipage}
\caption{Calculated polar diagrams for $\ket{\tilde{X}}$ (blue curves) and $\ket{\tilde{Y}}$ (red curves) emission intensity. The purple curve is the total emission intensity. The different parameters are: for $\theta = 0\degres$, (a) $\beta = 0$ (no VBM); (b) $\beta = 0.25$ and $\psi = 0\degres$; (c) $\beta = 0.25$ and $\psi = 45\degres$; (d) $\beta = 0.25$ and $\psi = 90\degres$, and for $\theta = +10\degres$, (e) $\beta = 0$ (no VBM); (f) $\beta = 0.25$ and $\psi = 0\degres$.}
\label{simu1}
\end{figure}

We performed simulations using equations (\ref{int1}), (\ref{int2}) and (\ref{int tot}) in order to understand the influence of the mixing parameters on the polarization properties of the exciton eigenstates. By plotting the total emission polar diagram (purple curve in Fig. \ref{psi0}), the mixing parameters $\beta$ and $\psi$ can be determined. The $\ket{\tilde{X}}$ and $\ket{\tilde{Y}}$ polar diagrams give also the QD main axis orientation angle $\theta$ (see Fig. \ref{simu1}). Indeed, as shown in (\ref{int tot}), the total intensity does not depend on $\theta$ and the parameter $\psi$ is given by the polar diagram orientation angle with respect to [1$\overline{1}$0]. $\beta$ is given by the contrast of the total PL polar diagram. The $\ket{\tilde{X}}$ and $\ket{\tilde{Y}}$ polar diagrams can then be fitted to determine the unknown parameter $\theta$. 

Calculated polar diagrams in Fig. \ref{simu1} show that when VBM occurs ($\beta \ne 0$), the dot total emission is partially linearly polarized along a direction defined by $\psi$, with a linear polarization degree $(I_{Max}-I_{min})/(I_{Max}+I_{min})$. In fact, this parameter depends only on $\beta$ and can be expressed as:
\begin{equation}
C(\beta)= \frac{2 \beta \sqrt{3(1-\beta^2)}}{3 - 2 \beta^2}
\label{contrast}
\end{equation}

\begin{figure}
\begin{center}
\subfigure[$\beta = 0.15$; $\psi = 10\degres$; $\theta = 0\degres$; $\phi_{\tilde{X}}=2\degres$; $\phi_{\tilde{Y}}=-1.5\degres$]{
      \includegraphics[width=0.3\textwidth]{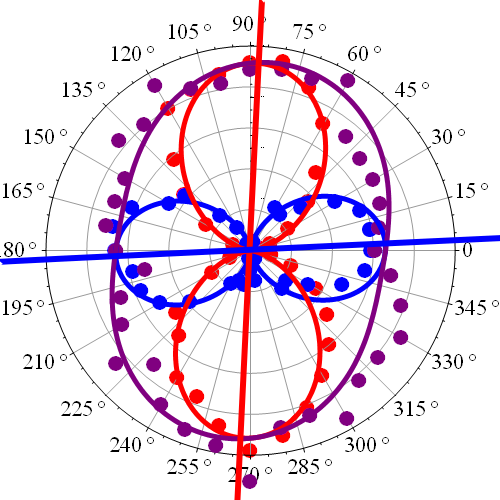}
\label{cahuetpouet1}}
\subfigure[$\beta = 0.15$; $\psi = 10\degres$; $\theta = -6\degres$; $\phi_{\tilde{X}}=9\degres$; $\phi_{\tilde{Y}}=3.5\degres$]{
      \includegraphics[width=0.3\textwidth]{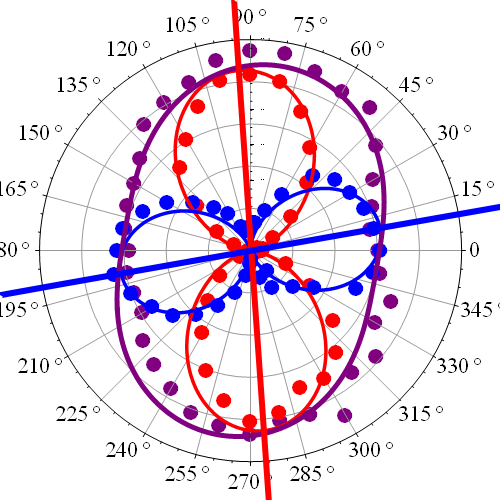}
\label{cahuetpouet2}}
\end{center}
\caption{Polar diagrams for $\ket{\tilde{X}}$ (bue dots) and $\ket{\tilde{Y}}$ (red dots) emission intensity, fitted by equations (\ref{int1}) (blue solid curve) and (\ref{int2}) (red solid curve) respectively. The total emission intensity is represented in purple. The two polar diagrams correspond to two different QDs located in the same detection spot with the same VBM parameters $\beta = 0.15$ and $\psi = 10\degres$, (a) $\theta = 0\degres$, $\lambda = 954.5nm$ and $\delta_{FSS} = 70\mu eV$; (b) $\theta = -6\degres$, $\lambda = 929.8nm$ and $\delta_{FSS} = 125\mu eV$.}
\label{cahuetpouetpouet}
\end{figure}

\begin{figure}
\begin{center}
\subfigure[]{
      \includegraphics[width=0.4\textwidth]{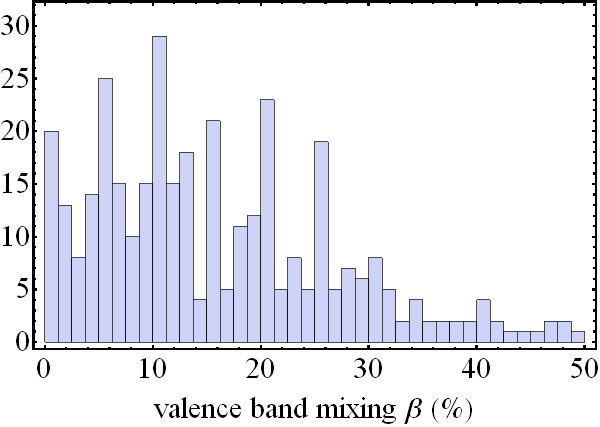}
\label{histobeta}}
\subfigure[]{
      \includegraphics[width=0.4\textwidth]{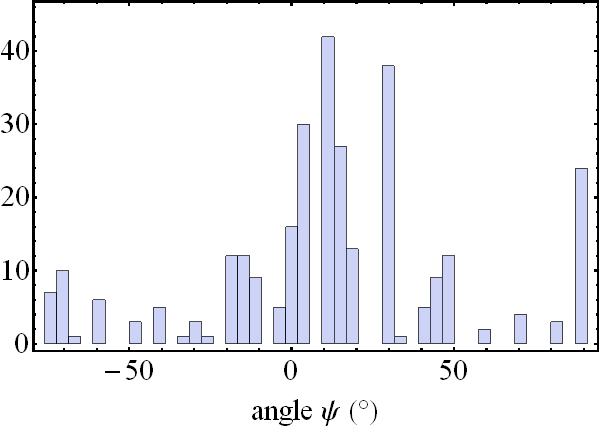}
\label{histopsi}}
\end{center}
\caption{Histograms of (a) $\beta$ and (b) $\psi$ in one of the studied samples. The emission wavelength of the QDs range between 900 nm and 940 nm.}
\label{histogr}
\end{figure}

Introducing, like in Refs. \citenum{Leger}. and \citenum{Lin}. the reduced variable $\eta = \frac{1}{\sqrt{3}} \frac{\beta}{\sqrt{1-\beta^2}}$, the linear polarization degree takes the simpler form: $C(\eta)= \frac{2 \eta}{1 + \eta^2}$. Depending on $\beta$ and $\psi$, the polarization of $\ket{\tilde{X}}$ (blue curve) and $\ket{\tilde{Y}}$ (red curve) emission can be tilted with respect to the crystallographic axes by an angle $\phi_{\tilde{X}}$ and $\phi_{\tilde{Y}}$ respectively. Moreover, the VBM allows for the $\ket{\tilde{X}}$ and $\ket{\tilde{Y}}$ states not to be orthogonal. This is shown for instance in Fig. \ref{simu1c},\ref{simu1d} and \ref{simu1h}. The tilt angles $\phi_{\tilde{X}}$ and $\phi_{\tilde{Y}}$ can be expressed as a function of $\eta$, $\theta$ and $\psi$ by:
\begin{equation}
\tan(\phi_{\tilde{X}})= \frac{-\sin\theta + \eta \sin(2\psi-\theta)}{\cos\theta - \eta \cos(2\psi-\theta)}
\label{phix}
\end{equation}
\begin{equation}
\tan(\phi_{\tilde{Y}})= - \frac{\sin\theta + \eta \sin(2\psi-\theta)}{\cos\theta + \eta \cos(2\psi-\theta)}
\label{phiy}
\end{equation}

First, let us consider the case where $\theta = 0\degres$ (see Fig. \ref{simu1a}, \ref{simu1b}, \ref{simu1c} and \ref{simu1d}), which occurs most often. The eigenstates tilting angles $\phi_{\tilde{X}}$ and $\phi_{\tilde{Y}}$ only depend on the mixing parameters $\psi$ and $\beta$ but in a non-trivial way. The general tendency is that for a fixed value of $\beta$, $\phi_{\tilde{X}}$ increases, reaches a maximum value, then decreases as a function of $\psi$. The larger $\beta$ is, the larger is the maximum value of $\phi_{\tilde{X}}$. $\phi_{\tilde{Y}}$ varies in an opposite way.

Now, let us consider the case $\theta\ne 0\degres$ (Fig. \ref{simu1g} and \ref{simu1h}). For a given $\psi$ the presence of a small tilting of the dot changes the eigenstates orientation and thus the angle $\Omega$ between them. For instance, we can see that for $\psi = 0\degres$, the two eigenstates are perpendicular when $\theta=0\degres$ ($\Omega=90\degres$, Fig. \ref{simu1b}), but when $\theta=10\degres$ the angle is $\Omega=96\degres$ (Fig. \ref{simu1h}). Thus, it appears that $\Omega$ does not depend only on $\psi$ and $\beta$ but depends also strongly on $\theta$.

Polarization resolved $\mu PL$ measurements were performed on a very large statistical ensemble of individual QDs. Non resonant excitation energy has been used, typically in the absorption region of the wetting layer around 1.4 eV. Fig. \ref{cahuetpouet1} and \ref{cahuetpouet2} show two typical experimental polar diagrams of two different QDs inside the same one $\mu m^2$ detection spot. For these two particular dots, the same values for $\beta$ and $\psi$ parameters were found ($\beta = 0.15$ and $\psi = 10\degres$) but the orientation angle $\theta$ was different ($\theta = 0\degres$ and $\theta = -6\degres$ for (a) and (b) respectively). Therefore, for a given total emission intensity (purple curve) it is not possible to determine the polarization direction of the two eigenstates (red and blue curves). In a general manner, we observed that all the dots (about 10 to 20) inside one $\mu m^2$ detection spot have the same value of angle $\psi$ but the VBM amplitude $\beta$ is very different for each dot, ranging between 0 and 0.5. In average, the mixing parameter is equal to 0.15 but with a large dispersion of $\pm 0.11$ (see Fig. \ref{histobeta}). The $\psi$ angle can also vary significantly depending on the dot position along the WG axis. However the $\psi$ values are mainly concentrated around $10\degres$ and $30\degres$ (Fig. \ref{histopsi}). Regarding the dots orientation axis $\theta$, rather small angles have been found, less than $15\degres$ and in most of the cases $\theta$ was smaller than $5\degres$. 

The fact that the angle $\psi$ is quite constant over a scale of one micron seems to indicate that $\psi$ is related mostly to the in-plane strain anisotropy which is the only parameter for which we can define a correlation length. As also described in Ref. \citenum{Leger}. $\psi$ is correlated to the main axis of the in-plane strain field, suggesting an homogeneous strain field inside the detection spot. The important variations found for $\beta$ on this same length scale indicate another physical origin having a strong influence on the mixing. Indeed, the VBM amplitude is related both to the strain and confinement potential anisotropy in the case of SK QDs. Since the in-plane strain anisotropy seems to be rather uniform over this distance, then the confinement potential should govern the amplitude of VBM. The non-uniformity of chemical composition, \cite{Lemaitre, Kegel} the differences in shape and size of the dots \cite{Joyce} can create important fluctuations of the confinement potential resulting in very dispersed values of the mixing parameter $\beta$. 

The strong VBM combined to the small misorientation of the dots results in tilted eigenstates with respect to the crystallographic axes. Tilt angles $\phi_{\tilde{X}}$ and $\phi_{\tilde{Y}}$ reaching $20\degres$, as well as an angle $\Omega$ ranging between $75\degres$ and $90\degres$ were found.

It is also worth noticing that in the case of the two dots presented in Fig. \ref{cahuetpouet1} and \ref{cahuetpouet2}, the $\beta$ parameter is the same but the emission wavelength and the FSS are different. Indeed, in a general manner we did not find any kind of correlation between the VBM amplitude and the emission wavelength or between the FSS and VBM, contrary to what has been recently reported in Ref. \citenum{Lin}. This disagreement might be due to the different confinement potential between the two kinds of samples, the emission energy of the dots being lower in energy in our case, suggesting a larger size. It is not also clear if the correlations reported in Ref. \citenum{Lin} are observed on a large area or in a very precise region on the sample. Our experimental results were obtained over hundreds of dots on different areas on the sample surface.

%%%%%%%%%%%%%%%%%%%%%%%%%%%%%%%%%%%%%%%%%%%%%%

%%%%\subsection{Under resonant excitation}

%%%%%%%%%%%%%%%%%%%%%%%%%%%%%%%%%%%%%%%%%%%%%%

\subsection{Resonant excitation}

\begin{figure}
\subfigure[$\beta = 0.3$; $\psi = 25\degres$; $\theta = 0\degres$; $R=0.7$]{
      \includegraphics[width=6cm,height=6cm]{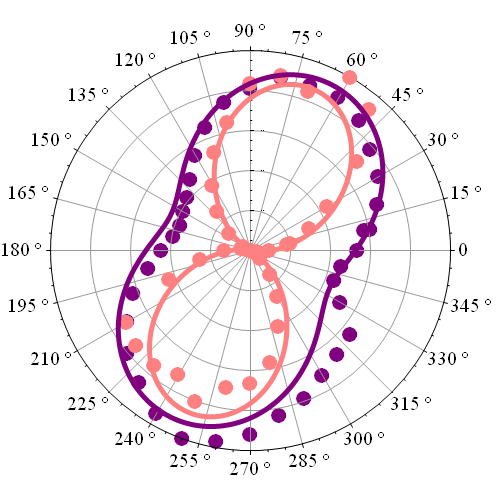}
\label{réso2}}
   \begin{minipage}[b]{0.2\linewidth}
\begin{center}
\subfigure[$R=3$]{
      \includegraphics[width=3.5cm,height=3.5cm]{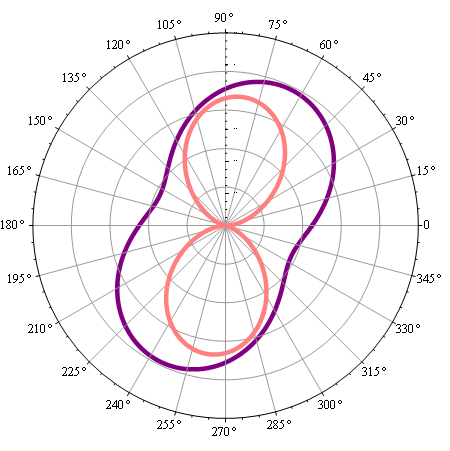}
\label{réso1}}
\end{center}
   \end{minipage} \hfill
   \begin{minipage}[b]{0.6\linewidth}
\begin{center}
\subfigure[$R=0.2$]{
      \includegraphics[width=3.5cm,height=3.5cm]{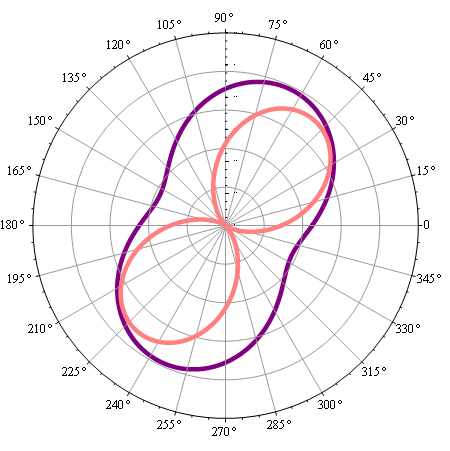}
\label{réso3}}
\end{center}
   \end{minipage}
\caption{Polar diagrams of the total emission intensity, for excitation off-resonance (purple curves) and on-resonance (pink curves). (a) represents experimental data for a given QD with $\beta = 0.3$, $\psi = 25\degres$, $\theta = 0\degres$ and $R = 0.7$. (b) and (c) are calculated plots with the same parameters $\beta$, $\psi$, $\theta$, and for an oscillator strength ratio $R =3$ and $R = 0.2$ respectively.}
\label{réso}
\end{figure}

In-plane PL polarization under resonant excitation has also been investigated to determine the oscillator strengths ratio $R = f_{\tilde{Y}}/f_{\tilde{X}}$ between the two $\ket{\tilde{Y}}$ and $\ket{\tilde{X}}$ eigenstates. Although the excitation laser is linearly polarized along the y-direction, both eigenstates can be excited since they can be tilted with respect to the crystallographic axes. Moreover the ps laser pulses are spectrally broader ($\approx 500 \mu eV$) than the FSS between the two eigenstates ($<150 \mu eV$). The pump power is kept low enough so the system is in the linear regime. Taking into account different oscillator strengths for the two transitions, and the proper laser polarization, the neutral exciton emission intensity is always linearly-polarized . The total resonant normalized intensity reads now:
\begin{eqnarray}
  \lefteqn{I_r(\alpha)  =  \{[\sqrt{1-\beta^2}(\cos \theta \sin \phi_{\tilde{X}} + \sin \theta \cos \phi_{\tilde{Y}})} \nonumber\\
  &&  -\frac{\beta}{\sqrt{3}} (\cos (\theta - 2 \psi) \sin \phi_{\tilde{X}}+ \sin( \theta - 2 \psi) \cos \phi_{\tilde{Y}})] \nonumber\\
  &&   (\cos \phi_{\tilde{X}} - R \sin \phi_{\tilde{Y}}) \cos \alpha \nonumber\\
  && +  [\sqrt{1-\beta^2}(-\sin \theta \sin \phi_{\tilde{X}} + \cos \theta \cos \phi_{\tilde{Y}}) \nonumber\\
  &&  -\frac{\beta}{\sqrt{3}} (\sin(\theta- 2 \psi) \sin \phi_{\tilde{X}}- \cos(\theta - 2 \psi) \cos \phi_{\tilde{Y}})]\nonumber\\
  &&  (\sin \phi_{\tilde{X}} + R \cos \phi_{\tilde{Y}}) \sin \alpha \}^2
\label{i réso}
\end{eqnarray}
$\phi_{\tilde{X}}$ and $\phi_{\tilde{Y}}$ can be replaced by expressions (\ref{phix}) and (\ref{phiy}), so $I_r(\alpha)$ depends only on $\beta$, $\psi$, $\theta$ and R. Using the VBM parameters $\beta$ and $\psi$ as well as the dot orientation angle $\theta$ that have been determined independently in non-resonant experiments, it is then possible for a given dot to obtain the ratio R. Fig. \ref{réso2} shows a typical polar diagram of the total emission intensity of a QD. The parameters $\beta=0.3$, $\psi=25\degres$ and $\theta=0\degres$ were found in off-resonance PL experiments and the ratio $R=0.7$ was given by the orientation of the resonant total intensity polar diagram (pink curve). Fig. \ref{réso1} and \ref{réso3} show how the value of the ratio R can change the total intensity orientation although the off-resonance parameters are the same. If the eigenstates were aligned with respect to the crystallographic axes ($\psi=0\degres$ or $90\degres$ and $\theta=0\degres$ or $90\degres$), then only the $\tilde{Y}$ state would be excited.

The oscillator strengths ratio was found to range between 0.2 ($f_{\tilde{X}}>f_{\tilde{Y}}$) and 1.5 ($f_{\tilde{X}} < f_{\tilde{Y}}$), but we noticed that in most cases, $f_{\tilde{X}}$ was larger than $f_{\tilde{Y}}$. This ratio can be significantly different from the ratio between the two eigenstates emission intensity found in off-resonance measurements. This fact confirms that in non resonant experiments the emission intensity is not proportional to the oscillator strength.\cite{Favero}

Finally, knowing the eigenstates polarization and oscillator strengths ratio for a given QD will allow addressing a particular state with the proper polarization and having a large oscillator strength. This will enhance the possibilities of coherent manipulations of an excitonic qu-bit. 

The in-plane PL polar diagrams under resonant excitation can also allow  to distinguish between the neutral exciton (linearly polarized) and the charged exciton (elliptically polarized). This can be useful in non-resonant experiments, because if the FSS can not be resolved both neutral and charged exciton lines appear to be elliptically polarized.

%%%%%%%%%%%%%%%%%%%%%%%%%%%%%%%%%%%%%%%%%%%%%%

%%%%%%%%%%%%%%%%%%%%%%%%%%%%%%%%%%%%%%%%%%%%%%

%%%%%\section{Valence-band mixing along the growth axis z}

%%%%%%%%%%%%%%%%%%%%%%%%%%%%%%%%%%%%%%%%%%%%%%

%%%%%%%%%%%%%%%%%%%%%%%%%%%%%%%%%%%%%%%%%%%%%%

\section{Valence-band mixing along the growth z-axis}

\begin{figure}[!t]
   \begin{minipage}[b]{0.2\linewidth}
\begin{center}
\subfigure[]{
      \includegraphics[width=5cm,height=5cm]{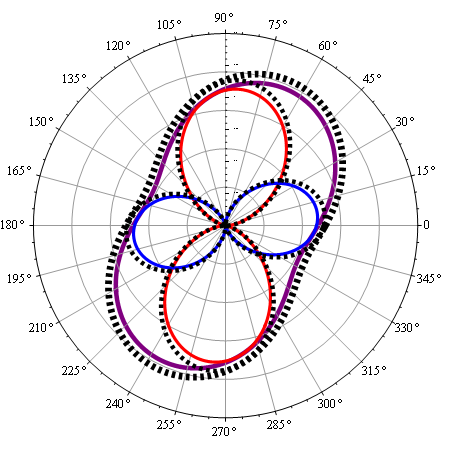}
\label{zbis}}
\end{center}
   \end{minipage} \hfill
   \begin{minipage}[b]{0.5\linewidth}
\begin{center}
\subfigure[]{
      \includegraphics[width=4cm,height=3cm]{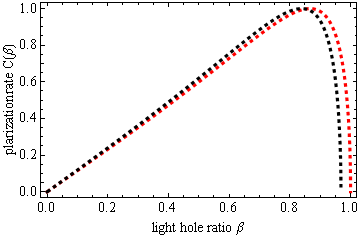}
\label{zter}}
\end{center}
   \end{minipage}
\caption{(a) Calculated in-plane polar diagrams under non resonant excitation with $\beta = 0.25$, $\psi= 25\degres$ and $\gamma = 0$ (red, blue and purple curves) and with $\gamma = 0.25$ (dashed black curve). (b) polarization rate $C(\beta, \gamma)$ versus the mixing parameter $\beta$ when $\gamma$ is fixed: $\gamma = 0$ (dashed red curve) and $\gamma = 0.25$ (dashed black curve).}
\label{zou}
\end{figure}

The original geometry of our set-up allows us, by exciting from the top surface and detecting the PL from the edge of the WG, to study the polarization properties along the z-axis, which is impossible in usual backscattering excitation-detection geometry. \cite{biref} Observation of PL with a polarization component along the z-axis in few dots, has led us to consider a mixing between heavy and light hole states with the same spins due to the different off-plane shear strain components which are usually neglected in the LKBP Hamiltonian. The eigenstates are written as linear combinations of the elliptically-polarized bright exciton states:
\begin{eqnarray}
  \lefteqn{\ket{\pm \tilde{1}}  =  \sqrt{1-\beta^2-\gamma^2} \ket{\mp \frac{1}{2}; \pm \frac{3}{2}}} \nonumber\\
& &+ \beta e^{\pm 2i\psi} \ket{\mp \frac{1}{2};\mp \frac{1}{2}} + \gamma e^{\pm 2i\xi}  \ket{\mp \frac{1}{2};\pm \frac{1}{2}}
\end{eqnarray}
where $\gamma$ represents opposite spin (in comparison to $\beta$) LH amplitude coupling with $\xi$ the associated phase. The existence of $\gamma$ has no significant influence on the polarization properties for in-plane PL of dots, and does not influence the determination of the three other parameters $\beta$, $\psi$ and $\theta$. Indeed, simulations (see Fig. \ref{zbis}) show that the presence of $\gamma$ changes neither the total emission intensity contrast (thus not changing $\beta$ and $\psi$), nor the two eigenstates emission intensity orientation, therefore not changing $\theta$. Only the normalization constants are slightly modified. We can also compare in Fig. \ref{zter} the polarization rates $C(\beta, \gamma)= \frac{2 \beta \sqrt{3(1-\beta^2-\gamma^2)}}{3 - 2 \beta^2-3\gamma^2}$ as a function of the mixing parameter $\beta$, for $\gamma=0$ (red curve) and  $\gamma=0.25$ (black curve). For the typical mixing strength we observed ($\beta< 0.5$), $C(\beta)$ is almost the same in both cases. We can also notice that $C(\beta)$ increases linearly until $\beta = 0.85$ and then decreases very fast. This is due to the fact that when $\beta$ tends to 1, we have pure LH states and the total emission intensity ($I_{\tilde{X}} + I_{\tilde{Y}}$) is then linearly polarized. The maximum of $C(\beta)$ does not occur for $\beta = 0.5$ because as we can see in appendix \ref{appendix1}, the expressions of $u_{\pm 3/2}^v$ and $u_{\pm 1/2}^v$ differ by a normalization constant ($1/\sqrt{2}$ and $1/\sqrt{6}$ respectively). On the contrary, the polarization degree in the (y,z) plane greatly depends on $\gamma$. The transitions dipole matrix elements taking into account the new VBM parameter $\gamma$, give the PL normalized intensity detected along the WG x-axis $I'_{nr}(\alpha)$, which reads:
\begin{eqnarray}
  \lefteqn{I'_{nr}(\alpha) = [ (1-\frac{2\beta^2}{3}} \nonumber\\
& & +2\beta \sqrt{\frac{1-\beta^2-\gamma^2}{3}} \cos(2\psi))  \sin^2 \alpha \nonumber\\
& & +\gamma^2(-\sin^2 \alpha+\frac{4}{3}\cos^2\alpha)]
\label{iz}
\end{eqnarray}

\begin{figure}[!t]
   \begin{minipage}[b]{0.2\linewidth}
\begin{center}
\subfigure[$\gamma < 0.05$]{
      \includegraphics[width=3.5cm,height=3.5cm]{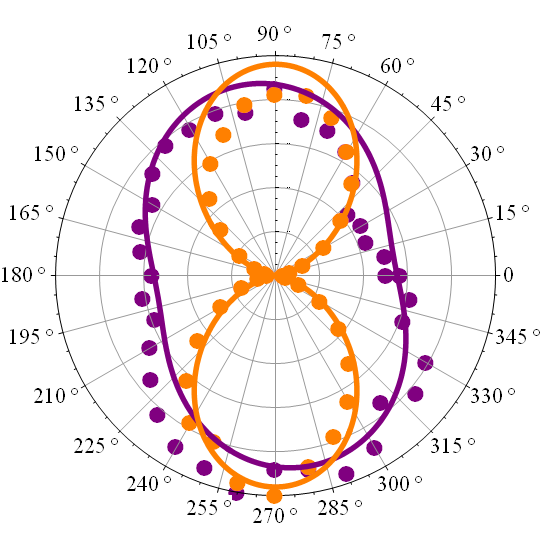}
\label{z1}}
\end{center}
   \end{minipage} \hfill
   \begin{minipage}[b]{0.6\linewidth}
\begin{center}
\subfigure[$\gamma = 0.25$]{
      \includegraphics[width=3.5cm,height=3.5cm]{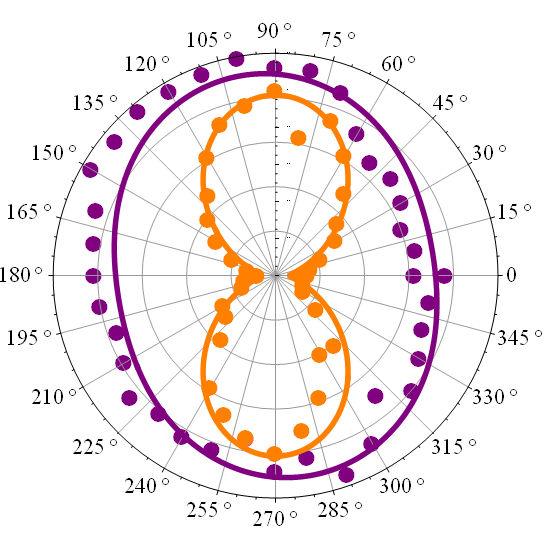}
\label{z2}}
\end{center}
   \end{minipage}
\caption{Experimental polar diagrams of two dots located in the same detection spot, under non resonant excitation, in the (x,y) plane (purple) and in the (y,z) plane (orange), fitted by equations (\ref{int tot}) and (\ref{iz}). (a) $\psi = - 25\degres$, $\beta = 0.2$ and $\gamma < 0.05$; (b) $\psi = - 25\degres$, $\beta = 0.09$ and $\gamma = 0.25$.}
\label{z}
\end{figure}

The mixing parameter $\gamma$ can be determined only if in-plane polarization resolved $\mu PL$ experiments have been previously performed on the same QD to determine $\beta$ and $\psi$. These combined experiments are difficult to perform because we have to observe the emission of the same QD detected from the top surface and from the edge of the WG. In most cases, $\gamma$ has a value too small ($\gamma<0.05$) to be accurately estimated (Fig. \ref{z1}), but for some dots $\gamma$ values range between $0.1$, and $0.25$ (Fig. \ref{z2}). Since $\gamma$ is related to the shear strain components $\epsilon_{zx}$ and $\epsilon_{yz}$, we can conclude that they modify also the dots optical properties and are partly responsible for the presence of emission polarized along the growth axis of the dots. 

%%%%%%%%%%%%%%%%%%%%%%%%%%%%%%%%%%%%%%%%%%%%%%

%%%%%%%%%%%%%%%%%%%%%%%%%%%%%%%%%%%%%%%%%%%%%%

%%%%\section{Conclusion}

%%%%%%%%%%%%%%%%%%%%%%%%%%%%%%%%%%%%%%%%%%%%%%

%%%%%%%%%%%%%%%%%%%%%%%%%%%%%%%%%%%%%%%%%%%%%%

\section{Conclusion}

In summary, the polarization-resolved $\mu PL$ experiments performed on single self-assembled InAs/GaAs QDs have allowed to determine the influence of the VBM on the polarization properties of the exciton eigenstates. Besides the strong polarization in-plane anisotropy, the most striking feature is that the linear states are neither aligned along the crystallographic axes nor orthogonal between them. The VBM strength can vary significantly depending on the dot position along the waveguide axis while the emission polarization axis has a given value within the detection spot size of one micron. Although strain and confinement anisotropy contributions to the VBM are difficult to separate, we can suggest that the strength of the mixing depends more strongly on confinement while the main polarization axis is mostly related to strain anisotropy.  Combining the in-plane polarization-resolved $\mu PL$ experiments with resonant and off-resonance excitation, the ratio of oscillator strengths has been determined. Knowing all the polarization properties and oscillator strength of the linear exciton states it is then possible to perform real coherent manipulation of single qu-bits and consider future possibilities for their entanglement.
The waveguiding geometry has also allowed observing emission polarized along the growth axis which is not possible to detect in usual backscattering photoluminescence experiments. This property is clearly related to shear components of the deformation tensor out of the growth plane. 
We did not find any relationship between the strength of the mixing or the tilt of the eigenstates polarization axis and the emission energy of the exciton. We believe that the correlation between emission energy and size of the dot can not be predicted in a straightforward manner especially when studying a large statistical ensemble of individual dots. Therefore it is difficult to establish general trends for the exciton fine structure versus the dots size unless an external field, like an electric field, \cite{Gong} or stress applied to the structure, \cite{Plumhof, Bennett} is used to tune the fine structure splitting.

%%%%%%%%%%%%%%%%%%%%%%%%%%%%%%%%%%%%%%%%%%%%%%

%%%%%%%%%%%%%%%%%%%%%%%%%%%%%%%%%%%%%%%%%%%%%%

%%%%\appendix

%%%%%%%%%%%%%%%%%%%%%%%%%%%%%%%%%%%%%%%%%%%%%%

%%%%%%%%%%%%%%%%%%%%%%%%%%%%%%%%%%%%%%%%%%%%%%

\appendix

\section{Wavefunctions}

We first assume that the band-gap energy, as well as the spin-orbital split-off energy are large enough to prevent any sizable coupling between the $\Gamma_6$, $\Gamma_7$ and $\Gamma_8$ bands, and we consider that the initial and final electron states are eigenstates separately described by the valence-band $\Gamma_8$ and the conduction-band $\Gamma_6$ effective-mass Hamiltonians \cite{Bastard}. In the envelope-function approximation, the conduction band $\Gamma_6$ and the valence-band $\Gamma_8$ wave functions are written as:
\begin{equation}
\psi_{ \pm 1/2}^c (\textbf{r}) = \chi_{e} (\textbf{r}) u_{ \pm 1/2}^c (\textbf{r})
\end{equation}
\begin{equation}
\psi_{ \pm 3/2}^v (\textbf{r}) = \chi_{hh} (\textbf{r}) u_{ \pm 3/2}^v (\textbf{r})
\end{equation}
\begin{equation}
\psi_{ \pm 1/2}^v (\textbf{r}) = \chi_{lh} (\textbf{r}) u_{ \pm 1/2}^v (\textbf{r})
\end{equation}
where $\chi_{e}$, $\chi_{hh}$ and $\chi_{lh}$ are the envelope-functions and $u_{ \pm 1/2}^c$, $u_{ \pm 1/2}^c$, $u_{ \pm 1/2}^c$ are the $\Gamma_6$ and $\Gamma_8$ Bloch functions, defined as \cite{Bastard}:
\begin{equation}
u_{1/2}^c = i \ket{s \uparrow}
\end{equation}
\begin{equation}
u_{-1/2}^c = i \ket{s \downarrow}
\end{equation}
\begin{equation}
u_{3/2}^v = \frac{1}{\sqrt{2}} \ket{(x+iy) \uparrow}
\end{equation}
\begin{equation}
u_{-1/2}^v = -\frac{1}{\sqrt{6}} \ket{(x-iy) \uparrow} - \sqrt{\frac{2}{3}} \ket{z \downarrow}
\end{equation}
\begin{equation}
u_{1/2}^v = \frac{1}{\sqrt{6}} \ket{(x+iy) \downarrow} - \sqrt{\frac{2}{3}} \ket{z \uparrow}
\end{equation}
\begin{equation}
u_{-3/2}^v = - \frac{1}{\sqrt{2}} \ket{(x-iy) \downarrow}
\end{equation}
where $\ket{s}$, $\ket{x}$, $\ket{y}$ and $\ket{z}$ are the orbital-functions with symmetry s, x, y and z, and $\ket{\uparrow \downarrow}$  are the spin components, quantized along the z-axis.
For simplicity, we will assume that the ratio between the overlap integrals of the electron-HH and electron-LH envelope-functions is equal, i.e. $\braket{\chi_{hh}|\chi_{e}}/\braket{\chi_{lh}|\chi_{e}} \approx 1$.

\label{appendix1}

\section{Hamiltionian with Valence band mixing}

We assume that the valence band states in a strained InAs/GaAs QD can be described by a (4x4) Hamiltonian based on  Luttinger-Kohn and Bir-Pikus theories which reads \cite{Testelin}:
\begin{equation}
\begin{pmatrix}
    P + Q + V &  R & - \sqrt{2} S & 0 \\
   R^\dagger &  P - Q + V & 0 &  \sqrt{2} S \\
  - \sqrt{2} S^\dagger  & 0 &  P - Q + V &  R \\
   0 &  \sqrt{2} S^\dagger &  R^\dagger &  P + Q + V
\end{pmatrix}
\end{equation}
in the $\{$ $u_{3/2}^v$; $u_{-1/2}^v$; $u_{1/2}^v$; $u_{-3/2}^v$ $\}$ basis.

V represents the confinement potential that usually can be partially separated because the confinement along the growth z-axis is more important than the in-plane confinement: $V=V_z(z)+V_{xy}(x,y)$. The other matrix elements are given as a sum of kinetic terms and strain counterparts:
\begin{eqnarray}
P & = & - \frac{\hbar^2}{2 m_0} \gamma_1 (k_x^2 + k_y^2 + k_z^2) \nonumber\\
& & +  a_v (\epsilon_{xx} + \epsilon_{yy} + \epsilon_{zz})
\end{eqnarray}
\begin{eqnarray}
Q & = & - \frac{\hbar^2}{2 m_0} \gamma_2 (k_x^2 + k_y^2 - 2 k_z^2) \nonumber\\
& & + \frac{ b_v}{2} (\epsilon_{xx} + \epsilon_{yy} -2 \epsilon_{zz})
\end{eqnarray}
\begin{eqnarray}
 R & = & - \frac{\hbar^2}{2 m_0} \sqrt{3} (\gamma_2 (k_x^2 - k_y^2) - 2i \gamma_3 k_x k_y) \nonumber\\
  & & + \frac{ \sqrt{3}}{2} b_v (\epsilon_{xx} - \epsilon_{yy}) - i d_v \epsilon_{xy}
\end{eqnarray}
\begin{eqnarray}
 S & = & - \frac{\hbar^2}{2 m_0} \sqrt{6} \gamma_3 (k_x - i k_y) k_z \nonumber\\
  & & + \frac{d_v}{\sqrt{2}} (\epsilon_{zx} - i \epsilon_{yz})
\end{eqnarray}
$\gamma_1$, $\gamma_2$ and $\gamma_3$ are the modified Luttinger parameters, $m_0$ is the free electron mass, $a_v$ is the valence-band hydrostatic deformation potential, $b_v$ and $d_v$ are the shear deformation potentials along the [001] and [111] axes, and $\epsilon_{ij}$ are the components of the deformation tensor.

Using the HH band as the origin of the energies in the valence-band, the above Hamiltonian can be rewritten according to Ref. \citenum{Leger} as:
\begin{equation}
\begin{pmatrix}
   0 & \rho_s e^{-2i \psi_0} & \sigma_s & 0 \\
   \rho_s e^{2i \psi_0} & \Delta_{lh} & 0 & \sigma_s \\
   \sigma_s & 0 & \Delta_{lh} & \rho_s e^{-2i \psi_0} \\
   0 & \sigma_s & \rho_s e^{2i \psi_0} & 0
\end{pmatrix}
\end{equation}
where $\Delta_{lh}$ is the HH-LH energy separation, $\rho_s$ is the in-plane (x,y) HH-LH mixing strength, $\sigma_s$ is the (y,z) plane HH-LH mixing strength, and $\psi$ the angle of in-plane anisotropy main axis with respect to [1$\overline{1}$0].\\

\label{appendix2}

\section{Optical transitions matrix elements}

Without VBM, the dipole matrix element of a transition between the conduction-band $\Gamma_6$ and the valence-band $\Gamma_8$ states can be written as:
\begin{eqnarray}
  \lefteqn{\braket{\psi_{\mp}^c | \textbf{$\epsilon$} . \textbf{p} | \psi_{\pm}^v} = \braket{\pm 1 | \textbf{$\epsilon$} . \textbf{p} | 0} } \nonumber\\
& & = \braket{\chi_{e} | \chi_{hh}} \braket{u_{\mp 1/2}^c | \textbf{$\epsilon$} . \textbf{p} | u_{\pm 3/2}^v}
\end{eqnarray}
where the polarization vector \textbf{$\epsilon$} defines the orientation of the electric field of the linearly polarized excitation laser and \textbf{p} is the electron linear momentum.
When VBM exists, the valence-band states become:
\begin{eqnarray}
  \lefteqn{\ket{\tilde{\psi}_{\pm}^v} = \sqrt{1-\beta^2-\gamma^2} \ket{\chi_{hh}} \ket{u_{\pm 3/2}^v} } \nonumber\\
& & = + \beta e^{\pm 2i \psi} \ket{\chi_{lh}} \ket{u_{\mp 1/2}^v}  \nonumber\\
& & + \gamma e^{\pm 2i \xi} \ket{\chi_{lh}} \ket{u_{\pm 1/2}^v}
\end{eqnarray}
and the dipole transition matrix element between conduction and valence states is modified and reads:
\begin{eqnarray}
  \lefteqn{\braket{\psi_{\mp}^c | \textbf{$\epsilon$} . \textbf{p} | \tilde{\psi}_{\pm}^v} = \braket{\pm \tilde{1} | \textbf{$\epsilon$} . \textbf{p} | 0} } \nonumber\\
& & = \braket{\chi_{e} | \chi_{hh}} \sqrt{1-\beta^2-\gamma^2} \braket{u_{\mp 1/2}^c | \textbf{$\epsilon$} . \textbf{p} | u_{\pm 3/2}^v}  \nonumber\\
& & + \braket{\chi_{e} | \chi_{lh}} \beta e^{2i\psi_0} \braket{u_{\mp 1/2}^c | \textbf{$\epsilon$} . \textbf{p} | u_{\mp 1/2}^v} \nonumber\\
& & + \braket{\chi_{e} | \chi_{lh}} \gamma \braket{u_{\mp 1/2}^c | \textbf{$\epsilon$} . \textbf{p} | u_{\pm 1/2}^v}
\end{eqnarray}
We assume that the overlaps $\braket{\chi_{hh}|\chi_{e}}/\braket{\chi_{lh}|\chi_{e}} \approx 1$, and we introduce the mixing parameters $\beta$ and $\gamma$, related to $\rho_S$, $\sigma_s$ and $\Delta_{lh}$ by: $\frac{\beta}{\sqrt{1-\beta^2-\gamma^2}} =  \frac{\rho_s}{\Delta_{lh}}$ and $\frac{\gamma}{\sqrt{1-\beta^2-\gamma^2}} =   \sigma_s$.

\label{appendix3}

%%%%%%%%%%%%%%%%%%%%%%%%%%%%%%%%%%%%%%%%%%%%%%

%%%%%%%%%%%%%%%%%%%%%%%%%%%%%%%%%%%%%%%%%%%%%%

%%%%%%%%%%%%%%%%%%%%%%%%%%%%%%%%%%%%%%%%%%%%%%

%%%%%%%%%%%%%%%%%%%%%%%%%%%%%%%%%%%%%%%%%%%%%%

%%%%%%%%%%%%%%%%%%%%%%%%%%%%%%%%%%%%%%%%%%%%%%


\begin{thebibliography}{10}

\bibitem{Imamoglu}
A. Imamoglu, D. D. Awschalom, G. Burkard, D. P. DiVincenzo,D. Loss, M. Sherwin, and A. Small,
Phys. Rev. Lett. \textbf{83}, 4204 (1999)

\bibitem{Xu}
X. Xu, Y Wu, B Sun,Q Huang, J Cheng, D. G. Steel,A. S. Bracker, D. Gammon, C. Emary, and L. J. Sham,
Phys. Rev. Lett.  \textbf{99}, 097401 (2007)

\bibitem{Ramsay}
A. J. Ramsay, S. J. Boyle, R. S. Kolodka, J. B. B. Oliveira, J. Skiba-Szymanska, H.Y. Liu, M. Hopkinson, A. M. Fox, and M. S. Skolnick,
Phys. Rev. Lett. \textbf{100}, 197401 (2008)

\bibitem{Gammon}
D. Gammon, E. S. Snow, B. V. Shanabrook, D. S. Katzer, and D. Park,
Phys. Rev. Lett. \textbf{76}, 3005 (1996)

\bibitem{Bayer}
M. Bayer, G. Ortner, O. Stern, A. Kuther, A. A. Gorbunov, A. Forchel, P. Hawrylak, S. Fafard,  K. Hinzer, T. L. Reinecke, S. N. Walck, J. P. Reithmaier, F. Klopf, and F. Schäfer,
Phys. Rev. B \textbf{65}, 195315 (2002)

\bibitem{Favero}
I. Favero, G. Cassabois, C. Voisin, C. Delalande, Ph. Roussignol, R. Ferreira, C. Couteau, J. P. Poizat, and J. M. Gérard,
Phys. Rev. B \textbf{71}, 233304 (2005)

\bibitem{Goupalov}
 S.V. Goupalov, E. L. Ivchenko, and A. V. Kavokin,
Superlattices and Microstruc. \textbf{23}, 1205 (1998)

\bibitem{Bester}
 G. Bester, S. Nair, and A. Zunger,
Phys. Rev. B \textbf{67}, 161306 (2003)

\bibitem{Karlsson10}
K. F. Karlsson, M. A. Dupertuis, D. Y. Oberli, E. Pelucchi, A. Rudra, P. O. Holtz, and E. Kapon,
Phys. Rev. B \textbf{81}, 161307 (R) (2010)

\bibitem{Flissikowski}
T. Flissikowski, A. Hundt, M. Lowisch, M. Rabe, and F. Henneberger,
Phys. Rev. Lett. \textbf{86}, 3172 (2001)

\bibitem{Astakhov}
G. V. Astakhov, T. Kiessling, A. V. Platonov, T. Slobodskyy, S. Mahapatra, W. Ossau, G. Schmidt, K. Brunner, and L. W. Molenkamp,
Phys. Rev. Lett. \textbf{96}, 027402 (2006)

\bibitem{Koudinov}
A. V. Koudinov,  I.A. Akimov, Y. G. Kusrayev, and F. Henneberger,
Phys. Rev. B \textbf{70}, 241305(R) (2004)

\bibitem{Leger}
Y. Leger, L. Besombes, L. Maingault, and H. Mariette,
Phys. Rev. B \textbf{76}, 045331 (2007)

\bibitem{Ohno}
S. Ohno, S. Adachi, R. Kaji, S. Muto and H. Sasakura,
Appl. Phys. Lett. \textbf{98}, 161912 (2011)

\bibitem{Belhadj}
T. Belhadj, T. Amand, A. Kunold, C.-M. Simon, T. Kuroda, M. Abbarchi, T. Mano, K. Sakoda, S. Kunz, X. Marie, and B. Urbaszek,
Appl. Phys. Lett. \textbf{97}, 051111 (2010)

\bibitem{Kowalik}
K. Kowalik, O. Krebs, A. Lema\^{\i}tre, J. A. Gaj, and P. Voisin,
Phys. Rev. B \textbf{77}, 161305(R) (2008)

\bibitem{Plumhof} 
J. D.Plumhof, V. Krapek, F. Ding, K. D. Jöns, R. Hafenbrak, P. Klenovsky, A. Herklotz, K. Dörr, P. Michler, A. Rastelli, and O. G. Schmidt, Phys. Rev. B \textbf{83}, 121302 (2011)

\bibitem{Lin}
C. H. Lin, W.T. You, H. Y. Chou, S.J. Cheng, S.D. Lin and W.H. Chang,
Phys. Rev. B \textbf{83}, 075317 (2011)

\bibitem{Testelin}
C. Testelin, F. Bernardot, B. Eble, M. Chamarro,
Phys. Rev. B \textbf{79}, 195440 (2009)

\bibitem{LK}
J. M. Luttinger and W. Kohn, Phys. Rev. \textbf{97}, 869 (1955)

\bibitem{BP}
G. L. Bir and G. E. Pikus, Sov. Phys. Solid State \textbf{3}, 2221 (1962)

\bibitem{Melet2}
R. Melet, V. Voliotis, R. Grousson, X. L. Wang, A. Lemaître, and A.Martinez,
Superlatice and Microstruct. \textbf{43}, 474 (2008)

\bibitem{Melet}
R. Melet, V. Voliotis, A. Enderlin, D. Roditchev, X. L. Wang, T. Guillet, and R. Grousson,  Phys. Rev. B \textbf{78}, 073301 (2008)

\bibitem{Enderlin}
A. Enderlin, M. Ravaro, V. Voliotis, R. Grousson, X. L. Wang,
Phys. Rev. B \textbf{80}, 085301 (2009)

\bibitem{Karlson}
K. F. Karlsson, V. Troncale, D. Y. Oberli, A. Malko, E. Pelucchi, A. Rudra, E. Kapon,
Appl. Phys. Lett. \textbf{89}, 251113 (2006)

\bibitem{Bastard}
U. Bockelmann, G. Bastard
Phys. Rev. B \textbf{45}, 1688 (1992)

\bibitem{Gong}
M Gong, W Zhang, G-C Guo, and L. He,
Phys. Rev. Lett. \textbf{106}, 227401 (2011)

\bibitem{Bennett}
A. J. Bennett, M. A. Pooley, R. M. Stevenson, M. B. Ward, R. B. Patel, A. Boyer de la Giroday, N. Sköld, I. Farrer, C. A. Nicoll, D. A. Ritchie, and A. J. Shields,
Nat. Phys. \textbf{6}, 947 (2010)

\bibitem{Lemaitre}
A. Lemaître, G. Patriarche, and F. Glas,
Appl. Phys. Lett. \textbf{85}, 3717 (2004)

\bibitem{Kegel}
I. Kegel, T. H. Metzger, A. Lorke, J. Peisl, J. Stangl, G. Bauer, J. M. García and P. M. Petroff,
Phys. Rev. Lett. \textbf{85}, 1694 (2000)

\bibitem{Joyce}
P. B. Joyce, T. J. Krzyzewski, G. R. Bell, T. S. Jones, S. Malik, D. Childs, and R. Murray,
Phys. Rev. Lett. \textbf{62}, 10891 (2000)

\bibitem{biref}
Because of the WG birefringence, care has to be taken to study QDs only near the edges of the sample. Thus, the polarization properties of the QDs emission are not altered by the intrinsic birefringence of the structure.

\end{thebibliography}
\end{document}